\documentclass[
showpacs,
floatfix,
aps,  
prl,
twocolumn,
superscriptaddress,
amssymb,
]{revtex4-1}

\usepackage{times}
\usepackage{wasysym}
\usepackage{amssymb}
\usepackage{latexsym}
\usepackage[dvips]{graphicx}
\usepackage{graphicx}
\usepackage{dcolumn}
\usepackage{amsfonts}
\usepackage{bm}
\usepackage{epsfig}

\newcommand{\be}{\begin{equation}}
\newcommand{\ee}{\end{equation}}
\newcommand{\bea}{\begin{eqnarray}}
\newcommand{\eea}{\end{eqnarray}}

\newcommand{\pc}{{\mathcal P}}

\def\Eac{\mathcal{E}_{\mathrm{ac}}}

\def\m{m^\star}
\def\eac{\epsilon}

\def\epseff{\epsilon_{\mathrm{eff}}}

\def\oc{\omega_{\mbox{\scriptsize {c}}}}

\def\tq{\tau_{\rm q}}
\def\ttr{\tau}
\def\tem{\tau_{\rm em}}

\def\tee{\tau_{\rm ee}}

\def\tqim{\tau_{\mbox{\scriptsize {q,0}}}}

\def\A{\mathcal{A}}

\newcommand{\req}[1]{Eq.\,(\ref{#1})}

\newcommand{\rfig}[1]{Fig.\,\ref{#1}}

\newcommand{\rref}[1]{Ref.\,\onlinecite{#1}}

\def\betao{\beta_{\omega}}
\def\A{\mathcal{A}}

\def\ne{n_{\rm e}}

\def\fc{{\small$\CIRCLE$}}

\def\fs{{\small$\blacksquare$}}

\def\naa{2.57}
\def\tqaa{44}

\def\tqab{23}
\def\ta{0.3}

\def\nba{3.33}
\def\tqba{32}

\def\tqbb{16}
\def\tb{1.8}

\begin{document}
\title{
Effect of illumination on quantum lifetime in GaAs quantum wells
}
\author{X. Fu}
\affiliation{School of Physics and Astronomy, University of Minnesota, Minneapolis, Minnesota 55455, USA}
\author{A. Riedl}
\affiliation{School of Physics and Astronomy, University of Minnesota, Minneapolis, Minnesota 55455, USA}
\author{M. Borisov}
\affiliation{School of Physics and Astronomy, University of Minnesota, Minneapolis, Minnesota 55455, USA}
\author{M. A. Zudov}
\email[Corresponding author: ]{zudov001@umn.edu}
\affiliation{School of Physics and Astronomy, University of Minnesota, Minneapolis, Minnesota 55455, USA}
\author{J. D. Watson$^\#$}
\affiliation{Department of Physics and Astronomy and Birck Nanotechnology Center, Purdue University, West Lafayette, Indiana 47907, USA}
\author{G. Gardner}
\affiliation{Department of Physics and Astronomy and Birck Nanotechnology Center, Purdue University, West Lafayette, Indiana 47907, USA}
\author{M. J. Manfra}
\affiliation{Department of Physics and Astronomy and Birck Nanotechnology Center, Purdue University, West Lafayette, Indiana 47907, USA}
\affiliation{Station Q Purdue, Purdue University, West Lafayette, Indiana 47907, USA}
\affiliation{School of Materials Engineering and School of Electrical and Computer Engineering, Purdue University, West Lafayette, Indiana 47907, USA}
\author{K. W. Baldwin}
\affiliation{Department of Electrical Engineering, Princeton University, Princeton, New Jersey 08544, USA}
\author{L. N. Pfeiffer}
\affiliation{Department of Electrical Engineering, Princeton University, Princeton, New Jersey 08544, USA}
\author{K. W. West}
\affiliation{Department of Electrical Engineering, Princeton University, Princeton, New Jersey 08544, USA}
\received{16 August 2018; revised manuscript received 7 October 2018; published 5 November 2018}

\begin{abstract}
Low-temperature illumination of a two-dimensional electron gas in GaAs quantum wells is known to greatly improve the quality of high-field magnetotransport.
The improvement is known to occur even when the carrier density and mobility remain unchanged, but what exactly causes it remains unclear. 
Here, we investigate the effect of illumination on microwave photoresistance in low magnetic fields. We find that the amplitude of microwave-induced resistance oscillations grows dramatically after  illumination. 
Dingle analysis reveals that this growth reflects a substantial increase in the single-particle (quantum) lifetime, which likely originates from the light-induced redistribution of charge enhancing the screening capability of the doping layers.
\end{abstract}

\maketitle

Even though low-temperature illumination of a two-dimensional electron gas (2DEG) in GaAs quantum wells is known to improve the quality of high-field magnetotransport \citep{cooper:2001,gamez:2013,samani:2014}, systematic investigations of this effect remain limited. 
One study \citep{gamez:2013} has investigated the effect of illumination on a 2DEG residing in a 30 nm-wide GaAs/Al$_{0.34}$Ga$_{0.66}$As quantum well with Si $\delta$-doping layers placed directly in Al$_{0.34}$Ga$_{0.66}$As barriers on both sides at setback distances of 100 nm (above the well) and 120 nm (below the well).
The initial effect of illumination is a considerable increase of both the density $\ne$ and the mobility $\mu$  of the 2DEG \citep{pfeiffer:1989} which, predictably, resulted in better developed fractional quantum Hall (FQH) states. 
However, additional, higher-intensity illumination left $\ne$ and $\mu$ essentially unchanged, while the transport features, e.g., the fragile FQH states in the $N=1$ Landau level, were further improved.
This improvement was attributed to the enhanced screening of ionized impurities by an increased number of polarized neutral shallow donors \citep{note:00}.

Another study  \citep{samani:2014} investigated the effect of illumination in a 2DEG hosted by a 30 nm-wide GaAs/Al$_{0.24}$Ga$_{0.76}$As quantum well utilizing a ``modern'' doping scheme.
This heterostructure was also remotely doped on both sides, but Si atoms were placed inside very narrow GaAs ``doping'' wells sandwiched between thin AlAs layers \citep{baba:1983,friedland:1996,pfeiffer:2003,umansky:2009,manfra:2014,gardner:2016,sammon:2018}.
Such doping scheme avoids formation of deep donor states, all Si atoms are ionized, but a significant fraction of donated electrons populate the X-band in surrounding AlAs layers.
Interestingly, illumination of such structure can also lead to improvement of high-field transport characteristics even though it does not appreciably change $\ne$ and $\mu$. 
For example, \rref{samani:2014} has shown that illumination can significantly enhance the measured energy gap of the FQH state at filling factor $\nu = 5/2$ and better development of other fragile quantum Hall states.
The enhancement of transport quality was linked to improved homogeneity of the 2DEG achieved after illumination.

In this Rapid Communication we (i) examine the effect of illumination on the quality of the \emph{low-field} magnetotransport under microwave irradiation and (ii) quantitatively assess the effect of illumination on total (quantum) lifetime $\tq$, which is a measure of electron-remote impurity scattering.
To measure $\tq$ we employ microwave-induced resistance oscillations (MIRO) \citep{zudov:2001a,ye:2001} which, in contrast to Shubnikov-de Haas oscillations \citep{qian:2017b}, are believed to be largely immune to macroscopic density fluctuations.
We find that after illumination MIRO become more pronounced while extending to lower magnetic fields. 
The Dingle analysis reveals that the observed improvement is a result of significant enhancement of quantum lifetime which increases by a factor of about two. 
This enhancement presents strong evidence that illumination results in reduced scattering from remote impurities, presumably due to light-induced redistribution of charge improving the screening capability of the doping layers.
Whether or not the increase of $\tq$ also contributes to the improvement of high-field transport \citep{cooper:2001,gamez:2013,samani:2014} remains an open question \citep{umansky:2009,qian:2017b}.

While we have investigated several samples with similar outcomes, here we present the results from two samples which exhibited almost no change in mobility due to illumination.
The 2DEG in sample A (B) resides in a GaAs quantum well of width 30 nm (24.9 nm) surrounded by Al$_{x}$Ga$_{1-x}$As barriers with $x = 0.24$ ($x = 0.28$).
Sample A (B) utilized Si doping in narrow GaAs doping wells surrounded by thin AlAs layers and positioned at a setback distance of 75 nm (80 nm) on both sides of the GaAs well hosting the 2DEG \citep{note:0}.
Both samples were $4\times 4$ mm squares with eight indium contacts fabricated at the corners and the midsides. 
When cooled in the dark, sample A (B) had the density $\ne \approx \naa \times 10^{11}$ cm$^{-2}$ ($\ne \approx \nba \times 10^{11}$ cm$^{-2}$).
Low-temperature mobility was estimated to be $\mu \approx 1.5 \times 10^7$ cm$^2$V$^{-1}$s$^{-1}$ in sample A and $\mu \approx 1.6 \times 10^7$ cm$^2$V$^{-1}$s$^{-1}$ in sample B.
Measurements were performed in Faraday configuration; microwave radiation was delivered to the sample immersed in liquid $^{3}$He inside a superconducting solenoid via a rectangular (WR-28) stainless steel waveguide with the magnetic field was applied perpendicular to the 2DEG.
The longitudinal resistance $R$ in sample A (B) was recorded using a standard low-frequency (a few Hz) four-terminal lock-in technique under continuous irradiation by microwaves of $f = 34$ GHz ($f = 64$ GHz) at a constant coolant temperature $T \approx 0.3$ K ($T \approx 1.8$ K).

Both sample A and sample B were illuminated by visible light (either green or white light-emitting diode) via the microwave waveguide at zero magnetic field for 10 minutes.
For sample A, we followed a procedure outlined in \rref{samani:2014}; illumination at base temperature ($T \approx 0.3$ K in our case) following up by an annealing step at $T \approx 2.5$ K for 15 minutes.
For sample B, we used ``conventional'' illumination temperature of $T\approx 5$ K after which the sample was cooled down in the dark.
After illumination procedure, the density of sample A (B) increased only by $\approx 4\times 10^{9}$ cm$^{-2}$ ($\approx 9\times 10^{9}$ cm$^{-2}$) while the mobilities remained essentially unchanged.
However, as we show next, both illumination protocols yielded substantial improvement of the quality of low-field magnetotransport, manifested by more pronounced MIRO, which we link to the enhancement of the quantum lifetime.

\begin{figure}[t]
\includegraphics{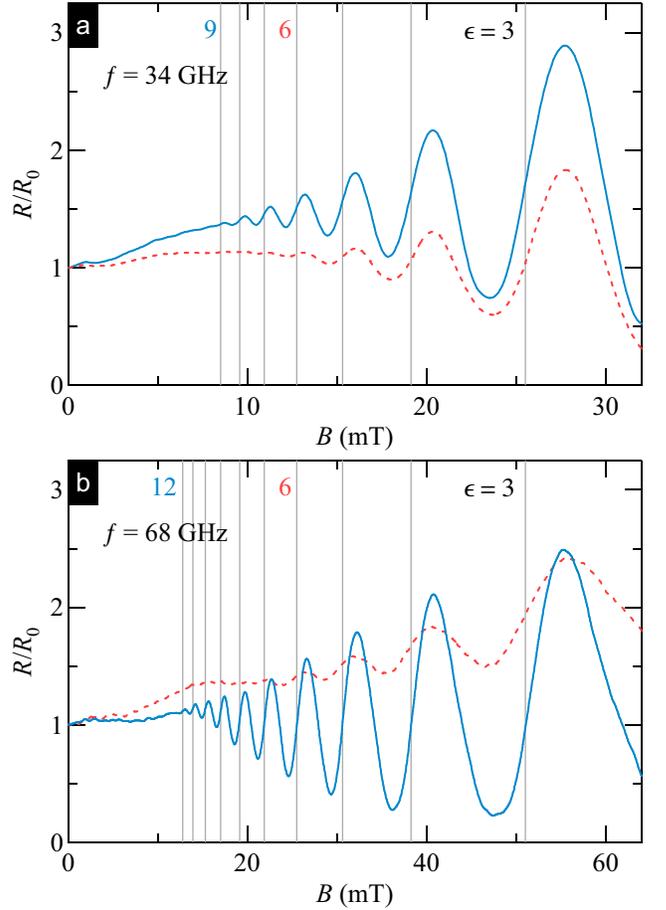}
\caption{(Color online)
Resistance in units of the zero-field resistance $R/R_0$ as a function of $B$ measured before (dotted line) and after (solid line) illumination in (a) sample A at $T \approx 0.3$ K and $f = 34$ GHz and (b) sample B at $T \approx 1.8$ K and $f = 68$ GHz.   
Vertical lines are drawn at integer $\eac$, as marked.
}
\label{fig1}
\end{figure}
Before presenting our experimental results, we recall that the oscillatory microwave photoresistance $\delta R$, i.e., the change of resistance caused by microwave radiation, can be written as \citep{dmitriev:2005,dmitriev:2009b,dmitriev:2012,note:cond}
\begin{equation}
\frac {\delta R (\eac)} {R_0} \propto -2\pi\eac\lambda^{2}\pc\sin2\pi\eac\,.
\label{eq.miro}
\end{equation}
Here, $R_0$ is the resistance at $B = 0$, $\eac=2\pi f /\oc$, $\oc = eB/\m$ is the cyclotron frequency, $\m \approx 0.06 m_0$ is the electron effective mass \citep{hatke:2013,shchepetilnikov:2017,fu:2017}, $\lambda = \exp(-\pi/\oc\tq)$ is the Dingle factor, and $\pc(\eac)$ is the effective microwave power which, for linearly polarized microwaves, is given by
\citep{chiu:1976,khodas:2008}
\be
\pc(\eac)= \frac {\pc^{0}} 2 \sum\limits_\pm\frac{1}{(1 \pm \eac^{-1})^2+\betao^2},~\pc^{0}=\frac{e^2\Eac^2v_F^2}{\epseff \hbar^2 \omega^4}\,,
\label{eq.pc}
\ee
where $\betao\equiv(\omega\tem)^{-1}+(\omega\ttr)^{-1}$, $\ttr = (\m/e)\mu$ is the momentum relaxation time, $\tem^{-1}=\ne e^2/2\sqrt{\epseff}\epsilon_0\m c$ \citep{chiu:1976,zhang:2014} is the radiative decay rate, $2\sqrt{\epseff}=\sqrt{\varepsilon}+1$ defines the effective dielectric constant $\epseff$, $\varepsilon=12.8$ is the dielectric constant of GaAs, $v_F$ is the Fermi velocity, and $\Eac$ is the microwave electric field.

The effect of low-temperature illumination on MIRO in sample A is illustrated in \rfig{fig1}(a) which shows the resistance $R$ normalized to its zero-field value $R_0$ measured before (dotted line) and after (solid line) illumination under microwave irradiation of frequency $f = 34$ GHz at temperature $T \approx \ta$ K.
Vertical lines are drawn at integer $\eac$, as marked.
The data clearly reveal that after illumination MIRO become more pronounced and extend to higher orders. 
Similar measurements in sample B, though employing different illumination procedure, yielded qualitatively identical results, as illustrated in \rfig{fig1}(b) showing the data at $f = 68$ GHz and $T \approx \tb$ K. 

\begin{figure}[t]
\includegraphics{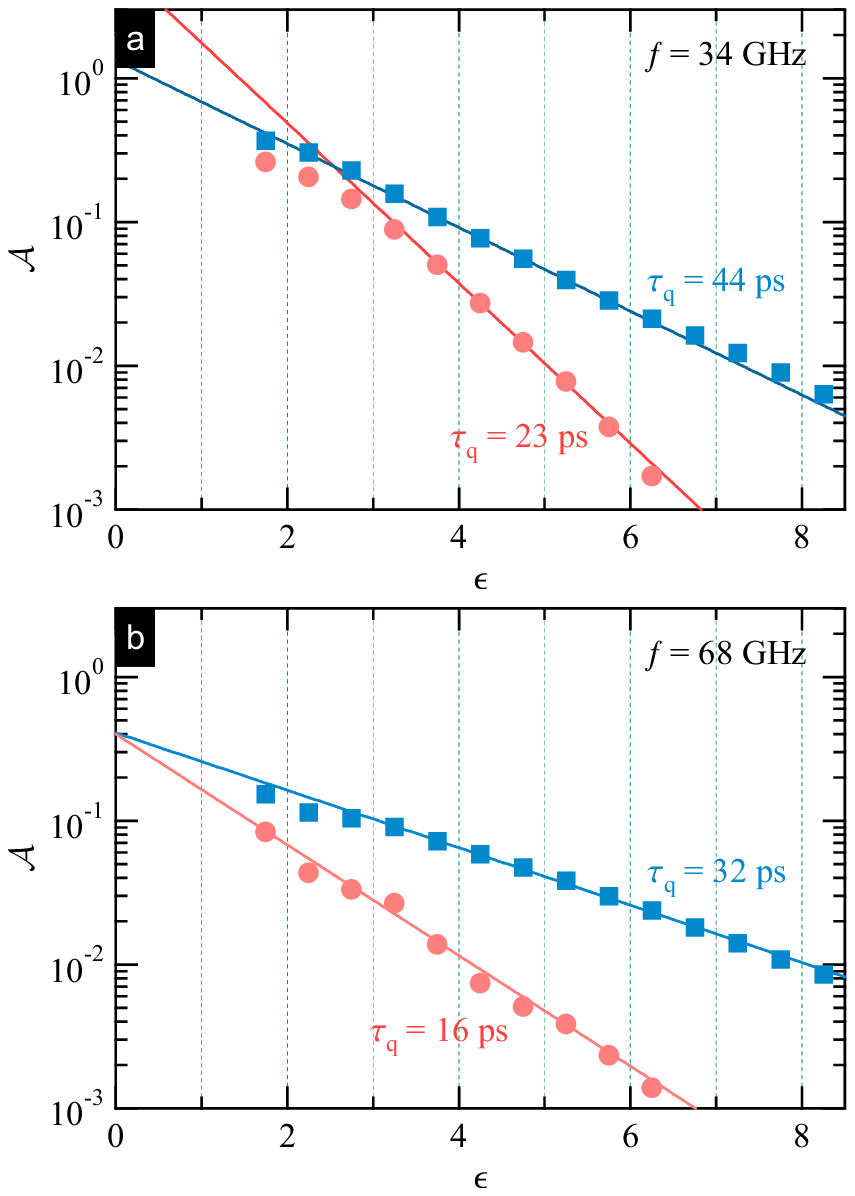}
\vspace{-0.1 in}
\caption{(Color online) 
Reduced MIRO amplitude $\A = |\delta R|_{\max}(\pc_0/\pc)/2\pi\eac R_0$ \citep{note:mass2} as a function of $\eac$ for (a) sample A and (b) sample B before (\fc) and after (\fs) illumination.
Fitting the data with $\A = \A_0\exp(-\eac/f\tq)$ (solid lines) reveals that illumination enhances $\tq$ from $\tqab$ ps to $\tqaa$ ps in sample A and from $\tqbb$ ps to $\tqba$ ps in sample B.
}
\vspace{-0.1 in}
\label{fig2}
\end{figure}

The results in \rfig{fig1} reveal that the enhancement of MIRO after illumination is significantly more pronounced at higher $\eac$, signaling an increase in quantum lifetime.
To quantify this increase, we performed Dingle analysis of MIRO.
Following \req{eq.miro}, we introduce a reduced MIRO amplitude $\A = |\delta R|_{\max}\pc_0/2\pi\eac\pc R_0$ \citep{note:mass2}, where $|\delta R|_{\max}$ is the measured MIRO amplitude.
The results for sample A and sample B are presented in \rfig{fig2}(a) and (b), respectively, which
show $\A$ as a function of $\eac$ extracted from the data acquired before (\fc) and after (\fs) illumination.
Fitting the data with $\A = \A_0\exp(-\eac/f\tq)$ (solid lines) reveals that illumination enhances $\tq$ from $\tqab$ ps to $\tqaa$ ps in sample A and from $\tqbb$ ps to $\tqba$ ps in sample B.

It is known that in contrast to Shubnikov-de Haas oscillations \citep{martin:2003,adamov:2006}, MIRO yield the quantum lifetime which is reduced by electron-electron scattering \citep{hatke:2009a}.
More specifically \citep{chaplik:1971,giuliani:1982,ryzhii:2003d,ryzhii:2004,dmitriev:2009b}, one can write: 
\be
\tq^{-1} = \tqim^{-1} + \tee^{-1}\,,
\label{eq.tq}
\ee
where $\tqim^{-1}$ represents the electron-impurity contribution and the electron-electron contribution $\tee^{-1}$ can be written as \citep{chaplik:1971,giuliani:1982,dmitriev:2005} 
\be
\frac\hbar\tee = \frac {\pi k_B^2 T^2} {4 E_F} \ln {\frac  {2 \hbar v_F/a_B}{\pi k_B T}}\,.
\label{eq.ee}
\ee
Here, $E_F$ is the Fermi energy and $a_B \approx 11$ nm is the Bohr radius in GaAs.
It is clear that subtracting the electron-electron contribution will only increase the change in impurity-limited quantum lifetime caused by illumination.
Because measurements on sample A were performed at low temperature ($T \approx 0.3$ K), electron-electron scattering rate is much smaller than $\tq^{-1} \approx \tqim^{-1}$.
In sample B, however, \req{eq.ee} yields $\tee \approx 80$ ps and using \req{eq.tq} we can estimate that $\tqim$ increases from $\tqim \approx 20$ ps to $\tqim \approx 53$ ps upon illumination. 

While the observed increase of quantum lifetime after illumination in both samples is quite significant, the momentum relaxation time $\ttr$ remained virtually unchanged.
This observation allows us to establish the source of disorder which is affected by illumination. 
Since the quantum scattering rate, in general, is much more sensitive to remote impurities than the transport scattering rate, we can conclude that illumination primarily affects scattering from remote impurities rather than from those in the vicinity of the GaAs quantum well.
Insensitivity of $\ttr$ to illumination then suggests that contribution of the remote impurities to the momentum relaxation rate is negligible even before the illumination, i.e., that $\ttr$ is limited by scattering from unintentional background impurities within the GaAs quantum well and in the AlGaAs barriers \citep{sammon:2018,sammon:2018a}.
The quantum scattering rate, on the other hand, can still contain a sizable or even dominant contribution from the remote impurities, e.g. Si ions in the doping layers, before the sample has been illuminated.

Recent theoretical examination \citep{sammon:2018,sammon:2018a} of the doping layers has shown that excess electrons which occupy the X-bands of the AlAs mini-wells form compact dipoles with donors of their choice (to minimize their energy) which reside in GaAs mini-wells.
These X-electrons can effectively screen the random potential from the remaining un-paired ionized Si atoms and the screening effectiveness grows rapidly with their number.
Because of this fast growth, the doping layer which has fewer X-electrons will contribute much more strongly to scattering than the other one.
In typical samples, such as ours, this would be the top doping layer which donate significant number of electrons to compensate surface states.
If the illumination can increase the number of X-electrons in the top doping layer, e.g., by returning electrons from the surface \citep{sammon:2018b}, one can expect a significant reduction of the quantum scattering rate.
Assuming that after illumination the number of X-electrons in the top doping layer becomes similar to that in the bottom doping layer, theoretical estimates \cite{sammon:2018,sammon:2018a} show that the remote impurity-limited quantum lifetime should be several times higher than observed in our experiment.
Our findings thus suggest that the quantum scattering rate after illumination is limited by scattering off background impurities residing in the main GaAs quantum well and in surrounding AlGaAs barriers.
 
While we clearly established that illumination significantly reduces quantum scattering rate, whether the observed reduction is the sole cause for the concurrent improvement in high-field transport characteristics \citep{cooper:2001,gamez:2013,samani:2014} can be debated \citep{umansky:2009,qian:2017b}.    
Indeed, as mentioned in \rref{samani:2014}, low-temperature illumination can also lead to improved density homogeneity of the 2DEG under study which must lead to improved development of FQH states, e.g., the increase of the excitation gap at $\nu = 5/2$.
MIRO, on the other hand, are nearly immune to macroscopic density fluctuations and therefore their enhancement can be linked directly to the increase of the quantum lifetime. 
Indeed, the enhancement of MIRO accompanied by an increase in $\tq$ has been observed even when samples became less homogeneous after illumination.

In summary, we have investigated the effect of low-temperature illumination on low-field magnetotransport characteristics of two-dimensional electrons in GaAs quantum wells subjected to microwave radiation.
We have found that microwave-induced resistance oscillations become significantly enhanced after illumination and that this enhancement is due to the increase of the quantum lifetime of the 2D electrons.
We believe that the observed increase likely originates from the light-induced redistribution of charge which increases the number of X-electrons in the top doping layer.
Insensitivity of transport scattering rate to illumination confirms that electron mobility is limited by background impurities in the vicinity of GaAs quantum well hosting the 2DEG even before illumination.

\begin{acknowledgements}
We thank M. Sammon and B. Shklovskii for discussions.
The work at Minnesota was supported by the NSF Award No. DMR-1309578.
Sample growth at Purdue was supported by the U.S. Department of Energy, Office of Science, Basic Energy Sciences, under Award DE-SC0006671.
L.N.P. and K.W.W. of Princeton University acknowledge the Gordon and Betty Moore Foundation
Grant No. GBMF 4420, and the National Science Foundation MRSEC Grant No. DMR-1420541.
\end{acknowledgements}

\small{$^\#$Present address: Microsoft Station-Q at Delft University of Technology, 2600 GA Delft, The Netherlands}


\begin{thebibliography}{37}
\expandafter\ifx\csname natexlab\endcsname\relax\def\natexlab#1{#1}\fi
\expandafter\ifx\csname bibnamefont\endcsname\relax
  \def\bibnamefont#1{#1}\fi
\expandafter\ifx\csname bibfnamefont\endcsname\relax
  \def\bibfnamefont#1{#1}\fi
\expandafter\ifx\csname citenamefont\endcsname\relax
  \def\citenamefont#1{#1}\fi
\expandafter\ifx\csname url\endcsname\relax
  \def\url#1{\texttt{#1}}\fi
\expandafter\ifx\csname urlprefix\endcsname\relax\def\urlprefix{URL }\fi
\providecommand{\bibinfo}[2]{#2}
\providecommand{\eprint}[2][]{\url{#2}}

\bibitem[{\citenamefont{Cooper et~al.}(2001)\citenamefont{Cooper, Lilly,
  Eisenstein, Jungwirth, Pfeiffer, and West}}]{cooper:2001}
\bibinfo{author}{\bibfnamefont{K.~B.} \bibnamefont{Cooper}},
  \bibinfo{author}{\bibfnamefont{M.~P.} \bibnamefont{Lilly}},
  \bibinfo{author}{\bibfnamefont{J.~P.} \bibnamefont{Eisenstein}},
  \bibinfo{author}{\bibfnamefont{T.}~\bibnamefont{Jungwirth}},
  \bibinfo{author}{\bibfnamefont{L.~N.} \bibnamefont{Pfeiffer}},
  \bibnamefont{and} \bibinfo{author}{\bibfnamefont{K.~W.} \bibnamefont{West}},
  \bibinfo{journal}{Solid State Commun.} \textbf{\bibinfo{volume}{119}},
  \bibinfo{pages}{89} (\bibinfo{year}{2001}).

\bibitem[{\citenamefont{Gamez and Muraki}(2013)}]{gamez:2013}
\bibinfo{author}{\bibfnamefont{G.}~\bibnamefont{Gamez}} \bibnamefont{and}
  \bibinfo{author}{\bibfnamefont{K.}~\bibnamefont{Muraki}},
  \bibinfo{journal}{Phys. Rev. B} \textbf{\bibinfo{volume}{88}},
  \bibinfo{pages}{075308} (\bibinfo{year}{2013}).

\bibitem[{\citenamefont{Samani et~al.}(2014)\citenamefont{Samani, Rossokhaty,
  Sajadi, L\"uscher, Folk, Watson, Gardner, and Manfra}}]{samani:2014}
\bibinfo{author}{\bibfnamefont{M.}~\bibnamefont{Samani}},
  \bibinfo{author}{\bibfnamefont{A.~V.} \bibnamefont{Rossokhaty}},
  \bibinfo{author}{\bibfnamefont{E.}~\bibnamefont{Sajadi}},
  \bibinfo{author}{\bibfnamefont{S.}~\bibnamefont{L\"uscher}},
  \bibinfo{author}{\bibfnamefont{J.~A.} \bibnamefont{Folk}},
  \bibinfo{author}{\bibfnamefont{J.~D.} \bibnamefont{Watson}},
  \bibinfo{author}{\bibfnamefont{G.~C.} \bibnamefont{Gardner}},
  \bibnamefont{and} \bibinfo{author}{\bibfnamefont{M.~J.}
  \bibnamefont{Manfra}}, \bibinfo{journal}{Phys. Rev. B}
  \textbf{\bibinfo{volume}{90}}, \bibinfo{pages}{121405}
  (\bibinfo{year}{2014}).

\bibitem[{\citenamefont{Pfeiffer et~al.}(1989)\citenamefont{Pfeiffer, West,
  Stormer, and Baldwin}}]{pfeiffer:1989}
\bibinfo{author}{\bibfnamefont{L.}~\bibnamefont{Pfeiffer}},
  \bibinfo{author}{\bibfnamefont{K.~W.} \bibnamefont{West}},
  \bibinfo{author}{\bibfnamefont{H.~L.} \bibnamefont{Stormer}},
  \bibnamefont{and} \bibinfo{author}{\bibfnamefont{K.~W.}
  \bibnamefont{Baldwin}}, \bibinfo{journal}{Appl. Phys. Lett.}
  \textbf{\bibinfo{volume}{55}}, \bibinfo{pages}{1888} (\bibinfo{year}{1989}).

\bibitem[{not({\natexlab{a}})}]{note:00}
\bibinfo{note}{Light-induced conversion of negatively-charged DX centers to
  neutral shallow donors has been demonstrated much earlier in
  \rref{hayne:1998} which systematically investigated the effect of
  illumination on time-dependent density and mobility in modulation-doped
  GaAs/Al$_{x}$Ga$_{1-x}$As single heterojunctions. This work also showed that
  illumination can increase mobility without affecting the carrier density.}

\bibitem[{\citenamefont{Baba et~al.}(1983)\citenamefont{Baba, Mizutani, and
  Ogawa}}]{baba:1983}
\bibinfo{author}{\bibfnamefont{T.}~\bibnamefont{Baba}},
  \bibinfo{author}{\bibfnamefont{T.}~\bibnamefont{Mizutani}}, \bibnamefont{and}
  \bibinfo{author}{\bibfnamefont{M.}~\bibnamefont{Ogawa}},
  \bibinfo{journal}{Jpn. J. of Appl. Phys.} \textbf{\bibinfo{volume}{22}},
  \bibinfo{pages}{L627} (\bibinfo{year}{1983}).

\bibitem[{\citenamefont{Friedland et~al.}(1996)\citenamefont{Friedland, Hey,
  Kostial, Klann, and Ploog}}]{friedland:1996}
\bibinfo{author}{\bibfnamefont{K.-J.} \bibnamefont{Friedland}},
  \bibinfo{author}{\bibfnamefont{R.}~\bibnamefont{Hey}},
  \bibinfo{author}{\bibfnamefont{H.}~\bibnamefont{Kostial}},
  \bibinfo{author}{\bibfnamefont{R.}~\bibnamefont{Klann}}, \bibnamefont{and}
  \bibinfo{author}{\bibfnamefont{K.}~\bibnamefont{Ploog}},
  \bibinfo{journal}{Phys. Rev. Lett.} \textbf{\bibinfo{volume}{77}},
  \bibinfo{pages}{4616} (\bibinfo{year}{1996}).

\bibitem[{\citenamefont{Pfeiffer and West}(2003)}]{pfeiffer:2003}
\bibinfo{author}{\bibfnamefont{L.}~\bibnamefont{Pfeiffer}} \bibnamefont{and}
  \bibinfo{author}{\bibfnamefont{K.~W.} \bibnamefont{West}},
  \bibinfo{journal}{Physica E} \textbf{\bibinfo{volume}{20}},
  \bibinfo{pages}{57} (\bibinfo{year}{2003}).

\bibitem[{\citenamefont{Umansky et~al.}(2009)\citenamefont{Umansky, Heiblum,
  Levinson, Smet, N{\"u}bler, and Dolev}}]{umansky:2009}
\bibinfo{author}{\bibfnamefont{V.}~\bibnamefont{Umansky}},
  \bibinfo{author}{\bibfnamefont{M.}~\bibnamefont{Heiblum}},
  \bibinfo{author}{\bibfnamefont{Y.}~\bibnamefont{Levinson}},
  \bibinfo{author}{\bibfnamefont{J.}~\bibnamefont{Smet}},
  \bibinfo{author}{\bibfnamefont{J.}~\bibnamefont{N{\"u}bler}},
  \bibnamefont{and} \bibinfo{author}{\bibfnamefont{M.}~\bibnamefont{Dolev}},
  \bibinfo{journal}{J. Cryst. Growth} \textbf{\bibinfo{volume}{311}},
  \bibinfo{pages}{1658} (\bibinfo{year}{2009}).

\bibitem[{\citenamefont{Manfra}(2014)}]{manfra:2014}
\bibinfo{author}{\bibfnamefont{M.~J.} \bibnamefont{Manfra}},
  \bibinfo{journal}{Annu. Rev. Condens. Matter Phys.}
  \textbf{\bibinfo{volume}{5}}, \bibinfo{pages}{347} (\bibinfo{year}{2014}).

\bibitem[{\citenamefont{Gardner et~al.}(2016)\citenamefont{Gardner, Fallahi,
  Watson, and Manfra}}]{gardner:2016}
\bibinfo{author}{\bibfnamefont{G.~C.} \bibnamefont{Gardner}},
  \bibinfo{author}{\bibfnamefont{S.}~\bibnamefont{Fallahi}},
  \bibinfo{author}{\bibfnamefont{J.~D.} \bibnamefont{Watson}},
  \bibnamefont{and} \bibinfo{author}{\bibfnamefont{M.~J.}
  \bibnamefont{Manfra}}, \bibinfo{journal}{J. Cryst. Growth}
  \textbf{\bibinfo{volume}{441}}, \bibinfo{pages}{71} (\bibinfo{year}{2016}).

\bibitem[{\citenamefont{Sammon et~al.}(2018{\natexlab{a}})\citenamefont{Sammon,
  Zudov, and Shklovskii}}]{sammon:2018}
\bibinfo{author}{\bibfnamefont{M.}~\bibnamefont{Sammon}},
  \bibinfo{author}{\bibfnamefont{M.~A.} \bibnamefont{Zudov}}, \bibnamefont{and}
  \bibinfo{author}{\bibfnamefont{B.~I.} \bibnamefont{Shklovskii}},
  \bibinfo{journal}{Phys. Rev. Materials} \textbf{\bibinfo{volume}{2}},
  \bibinfo{pages}{064604} (\bibinfo{year}{2018}{\natexlab{a}}).

\bibitem[{\citenamefont{Zudov et~al.}(2001)\citenamefont{Zudov, Du, Simmons,
  and Reno}}]{zudov:2001a}
\bibinfo{author}{\bibfnamefont{M.~A.} \bibnamefont{Zudov}},
  \bibinfo{author}{\bibfnamefont{R.~R.} \bibnamefont{Du}},
  \bibinfo{author}{\bibfnamefont{J.~A.} \bibnamefont{Simmons}},
  \bibnamefont{and} \bibinfo{author}{\bibfnamefont{J.~L.} \bibnamefont{Reno}},
  \bibinfo{journal}{Phys. Rev. B} \textbf{\bibinfo{volume}{64}},
  \bibinfo{pages}{201311(R)} (\bibinfo{year}{2001}).

\bibitem[{\citenamefont{Ye et~al.}(2001)\citenamefont{Ye, Engel, Tsui, Simmons,
  Wendt, Vawter, and Reno}}]{ye:2001}
\bibinfo{author}{\bibfnamefont{P.~D.} \bibnamefont{Ye}},
  \bibinfo{author}{\bibfnamefont{L.~W.} \bibnamefont{Engel}},
  \bibinfo{author}{\bibfnamefont{D.~C.} \bibnamefont{Tsui}},
  \bibinfo{author}{\bibfnamefont{J.~A.} \bibnamefont{Simmons}},
  \bibinfo{author}{\bibfnamefont{J.~R.} \bibnamefont{Wendt}},
  \bibinfo{author}{\bibfnamefont{G.~A.} \bibnamefont{Vawter}},
  \bibnamefont{and} \bibinfo{author}{\bibfnamefont{J.~L.} \bibnamefont{Reno}},
  \bibinfo{journal}{Appl. Phys. Lett.} \textbf{\bibinfo{volume}{79}},
  \bibinfo{pages}{2193} (\bibinfo{year}{2001}).

\bibitem[{\citenamefont{Qian et~al.}(2017)\citenamefont{Qian, Nakamura,
  Fallahi, Gardner, Watson, L\"uscher, Folk, Cs\'athy, and
  Manfra}}]{qian:2017b}
\bibinfo{author}{\bibfnamefont{Q.}~\bibnamefont{Qian}},
  \bibinfo{author}{\bibfnamefont{J.}~\bibnamefont{Nakamura}},
  \bibinfo{author}{\bibfnamefont{S.}~\bibnamefont{Fallahi}},
  \bibinfo{author}{\bibfnamefont{G.~C.} \bibnamefont{Gardner}},
  \bibinfo{author}{\bibfnamefont{J.~D.} \bibnamefont{Watson}},
  \bibinfo{author}{\bibfnamefont{S.}~\bibnamefont{L\"uscher}},
  \bibinfo{author}{\bibfnamefont{J.~A.} \bibnamefont{Folk}},
  \bibinfo{author}{\bibfnamefont{G.~A.} \bibnamefont{Cs\'athy}},
  \bibnamefont{and} \bibinfo{author}{\bibfnamefont{M.~J.}
  \bibnamefont{Manfra}}, \bibinfo{journal}{Phys. Rev. B}
  \textbf{\bibinfo{volume}{96}}, \bibinfo{pages}{035309}
  (\bibinfo{year}{2017}).

\bibitem[{not({\natexlab{b}})}]{note:0}
\bibinfo{note}{For a band diagram and a schematic of sample structure such as
  ours, see, e.g., Fig. 1 in \rref{umansky:2009} or Fig.\,7 in
  \rref{manfra:2014}. The schematic and discussion of the doping layer can also
  be found in \rref{sammon:2018}.}

\bibitem[{\citenamefont{Dmitriev et~al.}(2005)\citenamefont{Dmitriev, Vavilov,
  Aleiner, Mirlin, and Polyakov}}]{dmitriev:2005}
\bibinfo{author}{\bibfnamefont{I.~A.} \bibnamefont{Dmitriev}},
  \bibinfo{author}{\bibfnamefont{M.~G.} \bibnamefont{Vavilov}},
  \bibinfo{author}{\bibfnamefont{I.~L.} \bibnamefont{Aleiner}},
  \bibinfo{author}{\bibfnamefont{A.~D.} \bibnamefont{Mirlin}},
  \bibnamefont{and} \bibinfo{author}{\bibfnamefont{D.~G.}
  \bibnamefont{Polyakov}}, \bibinfo{journal}{Phys. Rev. B}
  \textbf{\bibinfo{volume}{71}}, \bibinfo{pages}{115316}
  (\bibinfo{year}{2005}).

\bibitem[{\citenamefont{Dmitriev et~al.}(2009)\citenamefont{Dmitriev, Khodas,
  Mirlin, Polyakov, and Vavilov}}]{dmitriev:2009b}
\bibinfo{author}{\bibfnamefont{I.~A.} \bibnamefont{Dmitriev}},
  \bibinfo{author}{\bibfnamefont{M.}~\bibnamefont{Khodas}},
  \bibinfo{author}{\bibfnamefont{A.~D.} \bibnamefont{Mirlin}},
  \bibinfo{author}{\bibfnamefont{D.~G.} \bibnamefont{Polyakov}},
  \bibnamefont{and} \bibinfo{author}{\bibfnamefont{M.~G.}
  \bibnamefont{Vavilov}}, \bibinfo{journal}{Phys. Rev. B}
  \textbf{\bibinfo{volume}{80}}, \bibinfo{pages}{165327}
  (\bibinfo{year}{2009}).

\bibitem[{\citenamefont{Dmitriev et~al.}(2012)\citenamefont{Dmitriev, Mirlin,
  Polyakov, and Zudov}}]{dmitriev:2012}
\bibinfo{author}{\bibfnamefont{I.~A.} \bibnamefont{Dmitriev}},
  \bibinfo{author}{\bibfnamefont{A.~D.} \bibnamefont{Mirlin}},
  \bibinfo{author}{\bibfnamefont{D.~G.} \bibnamefont{Polyakov}},
  \bibnamefont{and} \bibinfo{author}{\bibfnamefont{M.~A.} \bibnamefont{Zudov}},
  \bibinfo{journal}{Rev. Mod. Phys.} \textbf{\bibinfo{volume}{84}},
  \bibinfo{pages}{1709} (\bibinfo{year}{2012}).

\bibitem[{not({\natexlab{c}})}]{note:cond}
\bibinfo{note}{\req{eq.miro} was obtained assuming $2\pi k_B T \gg \hbar
  \omega$ and is accurate away from the cyclotron resonance ($2\pi\eac \gg 1$),
  when the microwave power is not too high ($\pc \ll 1$), and when Landau
  levels are overlapping ($\lambda \ll 1$). All these conditions are satisfied
  reasonably well in our experiment.}

\bibitem[{\citenamefont{Hatke et~al.}(2013)\citenamefont{Hatke, Zudov, Watson,
  Manfra, Pfeiffer, and West}}]{hatke:2013}
\bibinfo{author}{\bibfnamefont{A.~T.} \bibnamefont{Hatke}},
  \bibinfo{author}{\bibfnamefont{M.~A.} \bibnamefont{Zudov}},
  \bibinfo{author}{\bibfnamefont{J.~D.} \bibnamefont{Watson}},
  \bibinfo{author}{\bibfnamefont{M.~J.} \bibnamefont{Manfra}},
  \bibinfo{author}{\bibfnamefont{L.~N.} \bibnamefont{Pfeiffer}},
  \bibnamefont{and} \bibinfo{author}{\bibfnamefont{K.~W.} \bibnamefont{West}},
  \bibinfo{journal}{Phys. Rev. B} \textbf{\bibinfo{volume}{87}},
  \bibinfo{pages}{161307(R)} (\bibinfo{year}{2013}).

\bibitem[{\citenamefont{Shchepetilnikov
  et~al.}(2017)\citenamefont{Shchepetilnikov, Frolov, Nefyodov, Kukushkin, and
  Schmult}}]{shchepetilnikov:2017}
\bibinfo{author}{\bibfnamefont{A.~V.} \bibnamefont{Shchepetilnikov}},
  \bibinfo{author}{\bibfnamefont{D.~D.} \bibnamefont{Frolov}},
  \bibinfo{author}{\bibfnamefont{Y.~A.} \bibnamefont{Nefyodov}},
  \bibinfo{author}{\bibfnamefont{I.~V.} \bibnamefont{Kukushkin}},
  \bibnamefont{and} \bibinfo{author}{\bibfnamefont{S.}~\bibnamefont{Schmult}},
  \bibinfo{journal}{Phys. Rev. B} \textbf{\bibinfo{volume}{95}},
  \bibinfo{pages}{161305} (\bibinfo{year}{2017}).

\bibitem[{\citenamefont{Fu et~al.}(2017)\citenamefont{Fu, Ebner, Shi, Zudov,
  Qian, Watson, and Manfra}}]{fu:2017}
\bibinfo{author}{\bibfnamefont{X.}~\bibnamefont{Fu}},
  \bibinfo{author}{\bibfnamefont{Q.~A.} \bibnamefont{Ebner}},
  \bibinfo{author}{\bibfnamefont{Q.}~\bibnamefont{Shi}},
  \bibinfo{author}{\bibfnamefont{M.~A.} \bibnamefont{Zudov}},
  \bibinfo{author}{\bibfnamefont{Q.}~\bibnamefont{Qian}},
  \bibinfo{author}{\bibfnamefont{J.~D.} \bibnamefont{Watson}},
  \bibnamefont{and} \bibinfo{author}{\bibfnamefont{M.~J.}
  \bibnamefont{Manfra}}, \bibinfo{journal}{Phys. Rev. B}
  \textbf{\bibinfo{volume}{95}}, \bibinfo{pages}{235415}
  (\bibinfo{year}{2017}).

\bibitem[{\citenamefont{Chiu et~al.}(1976)\citenamefont{Chiu, Lee, and
  Quinn}}]{chiu:1976}
\bibinfo{author}{\bibfnamefont{K.~W.} \bibnamefont{Chiu}},
  \bibinfo{author}{\bibfnamefont{T.~K.} \bibnamefont{Lee}}, \bibnamefont{and}
  \bibinfo{author}{\bibfnamefont{J.~J.} \bibnamefont{Quinn}},
  \bibinfo{journal}{Surf. Sci.} \textbf{\bibinfo{volume}{58}},
  \bibinfo{pages}{182} (\bibinfo{year}{1976}).

\bibitem[{\citenamefont{Khodas and Vavilov}(2008)}]{khodas:2008}
\bibinfo{author}{\bibfnamefont{M.}~\bibnamefont{Khodas}} \bibnamefont{and}
  \bibinfo{author}{\bibfnamefont{M.~G.} \bibnamefont{Vavilov}},
  \bibinfo{journal}{Phys. Rev. B} \textbf{\bibinfo{volume}{78}},
  \bibinfo{pages}{245319} (\bibinfo{year}{2008}).

\bibitem[{\citenamefont{Zhang et~al.}(2014)\citenamefont{Zhang, Arikawa, Kato,
  Reno, Pan, Watson, Manfra, Zudov, Tokman, Erukhimova et~al.}}]{zhang:2014}
\bibinfo{author}{\bibfnamefont{Q.}~\bibnamefont{Zhang}},
  \bibinfo{author}{\bibfnamefont{T.}~\bibnamefont{Arikawa}},
  \bibinfo{author}{\bibfnamefont{E.}~\bibnamefont{Kato}},
  \bibinfo{author}{\bibfnamefont{J.~L.} \bibnamefont{Reno}},
  \bibinfo{author}{\bibfnamefont{W.}~\bibnamefont{Pan}},
  \bibinfo{author}{\bibfnamefont{J.~D.} \bibnamefont{Watson}},
  \bibinfo{author}{\bibfnamefont{M.~J.} \bibnamefont{Manfra}},
  \bibinfo{author}{\bibfnamefont{M.~A.} \bibnamefont{Zudov}},
  \bibinfo{author}{\bibfnamefont{M.}~\bibnamefont{Tokman}},
  \bibinfo{author}{\bibfnamefont{M.}~\bibnamefont{Erukhimova}},
  \bibnamefont{et~al.}, \bibinfo{journal}{Phys. Rev. Lett.}
  \textbf{\bibinfo{volume}{113}}, \bibinfo{pages}{047601}
  (\bibinfo{year}{2014}).

\bibitem[{not({\natexlab{d}})}]{note:mass2}
\bibinfo{note}{$\pc/\pc_0$ was calculated using $\m = 0.067m_0$ entering $\eac$
  and $\tem$.}

\bibitem[{\citenamefont{Martin et~al.}(2003)\citenamefont{Martin, Maslov, and
  Reizer}}]{martin:2003}
\bibinfo{author}{\bibfnamefont{G.~W.} \bibnamefont{Martin}},
  \bibinfo{author}{\bibfnamefont{D.~L.} \bibnamefont{Maslov}},
  \bibnamefont{and} \bibinfo{author}{\bibfnamefont{M.~Y.}
  \bibnamefont{Reizer}}, \bibinfo{journal}{Phys. Rev. B}
  \textbf{\bibinfo{volume}{68}}, \bibinfo{pages}{241309}
  (\bibinfo{year}{2003}).

\bibitem[{\citenamefont{Adamov et~al.}(2006)\citenamefont{Adamov, Gornyi, and
  Mirlin}}]{adamov:2006}
\bibinfo{author}{\bibfnamefont{Y.}~\bibnamefont{Adamov}},
  \bibinfo{author}{\bibfnamefont{I.~V.} \bibnamefont{Gornyi}},
  \bibnamefont{and} \bibinfo{author}{\bibfnamefont{A.~D.}
  \bibnamefont{Mirlin}}, \bibinfo{journal}{Phys. Rev. B}
  \textbf{\bibinfo{volume}{73}}, \bibinfo{pages}{045426}
  (\bibinfo{year}{2006}).

\bibitem[{\citenamefont{Hatke et~al.}(2009)\citenamefont{Hatke, Zudov,
  Pfeiffer, and West}}]{hatke:2009a}
\bibinfo{author}{\bibfnamefont{A.~T.} \bibnamefont{Hatke}},
  \bibinfo{author}{\bibfnamefont{M.~A.} \bibnamefont{Zudov}},
  \bibinfo{author}{\bibfnamefont{L.~N.} \bibnamefont{Pfeiffer}},
  \bibnamefont{and} \bibinfo{author}{\bibfnamefont{K.~W.} \bibnamefont{West}},
  \bibinfo{journal}{Phys. Rev. Lett.} \textbf{\bibinfo{volume}{102}},
  \bibinfo{pages}{066804} (\bibinfo{year}{2009}).

\bibitem[{\citenamefont{Chaplik}(1971)}]{chaplik:1971}
\bibinfo{author}{\bibfnamefont{A.~V.} \bibnamefont{Chaplik}},
  \bibinfo{journal}{Sov. Phys. JETP} \textbf{\bibinfo{volume}{33}},
  \bibinfo{pages}{997} (\bibinfo{year}{1971}).

\bibitem[{\citenamefont{Giuliani and Quinn}(1982)}]{giuliani:1982}
\bibinfo{author}{\bibfnamefont{G.~F.} \bibnamefont{Giuliani}} \bibnamefont{and}
  \bibinfo{author}{\bibfnamefont{J.~J.} \bibnamefont{Quinn}},
  \bibinfo{journal}{Phys. Rev. B} \textbf{\bibinfo{volume}{26}},
  \bibinfo{pages}{4421} (\bibinfo{year}{1982}).

\bibitem[{\citenamefont{Ryzhii and Suris}(2003)}]{ryzhii:2003d}
\bibinfo{author}{\bibfnamefont{V.}~\bibnamefont{Ryzhii}} \bibnamefont{and}
  \bibinfo{author}{\bibfnamefont{R.}~\bibnamefont{Suris}}, \bibinfo{journal}{J.
  Phys.: Condens. Matter} \textbf{\bibinfo{volume}{15}}, \bibinfo{pages}{6855}
  (\bibinfo{year}{2003}).

\bibitem[{\citenamefont{Ryzhii et~al.}(2004)\citenamefont{Ryzhii, Chaplik, and
  Suris}}]{ryzhii:2004}
\bibinfo{author}{\bibfnamefont{V.}~\bibnamefont{Ryzhii}},
  \bibinfo{author}{\bibfnamefont{A.}~\bibnamefont{Chaplik}}, \bibnamefont{and}
  \bibinfo{author}{\bibfnamefont{R.}~\bibnamefont{Suris}},
  \bibinfo{journal}{JETP Lett.} \textbf{\bibinfo{volume}{80}},
  \bibinfo{pages}{363} (\bibinfo{year}{2004}).

\bibitem[{\citenamefont{Sammon et~al.}(2018{\natexlab{b}})\citenamefont{Sammon,
  Tianran, and Shklovskii}}]{sammon:2018a}
\bibinfo{author}{\bibfnamefont{M.}~\bibnamefont{Sammon}},
  \bibinfo{author}{\bibfnamefont{C.}~\bibnamefont{Tianran}}, \bibnamefont{and}
  \bibinfo{author}{\bibfnamefont{B.~I.} \bibnamefont{Shklovskii}},
  \bibinfo{journal}{Phys. Rev. Materials} \textbf{\bibinfo{volume}{2}},
  \bibinfo{pages}{in press} (\bibinfo{year}{2018}{\natexlab{b}}).

\bibitem[{\citenamefont{Sammon and Shklovskii}(2018)}]{sammon:2018b}
\bibinfo{author}{\bibfnamefont{M.}~\bibnamefont{Sammon}} \bibnamefont{and}
  \bibinfo{author}{\bibfnamefont{B.~I.} \bibnamefont{Shklovskii}},
  \bibinfo{journal}{manuscript in preparation}  (\bibinfo{year}{2018}).

\bibitem[{\citenamefont{Hayne et~al.}(1998)\citenamefont{Hayne, Usher, Harris,
  Moshchalkov, and Foxon}}]{hayne:1998}
\bibinfo{author}{\bibfnamefont{M.}~\bibnamefont{Hayne}},
  \bibinfo{author}{\bibfnamefont{A.}~\bibnamefont{Usher}},
  \bibinfo{author}{\bibfnamefont{J.~J.} \bibnamefont{Harris}},
  \bibinfo{author}{\bibfnamefont{V.~V.} \bibnamefont{Moshchalkov}},
  \bibnamefont{and} \bibinfo{author}{\bibfnamefont{C.~T.} \bibnamefont{Foxon}},
  \bibinfo{journal}{Phys. Rev. B} \textbf{\bibinfo{volume}{57}},
  \bibinfo{pages}{14813} (\bibinfo{year}{1998}).

\end{thebibliography}

\end{document}